\documentclass[a4paper,11pt]{article}
\usepackage{pos}

\usepackage{graphicx}   
\usepackage{subcaption}
\usepackage{natbib}
\usepackage{siunitx}
\DeclareSIUnit{\baud}{Baud} 
\DeclareSIUnit[number-unit-product = {}]{\inchQ}{\textquotedbl}
\DeclareSIUnit[number-unit-product = {}]{\inch}{''}
\usepackage{cleveref}

\title{Electronics Design of the IceCube-Gen2 Optical Module Prototype}
 \ShortTitle{Electronics Design of the IceCube-Gen2 Optical Module Prototype}

\author{The IceCube-Gen2 Collaboration \\{\normalsize \normalfont(a complete list of authors can be found at the end of the proceedings)}\\}

\emailAdd{sgriffin@icecube.wisc.edu}

\abstract{

IceCube-Gen2 is a planned extension to the existing IceCube Neutrino Observatory and will provide an order of magnitude increase in the detection rate of cosmic neutrinos by deploying ~10,000 sensors in a volume of ~8 cubic kilometers. 
As part of the upcoming IceCube Upgrade, we are developing prototype IceCube-Gen2 sensors to test all components in-situ in preparation for mass production required for IceCube-Gen2. The novel IceCube-Gen2 module will contain up to eighteen 4-inch photomultiplier tubes (PMTs). 
The signals for each PMT are digitized with a 2-channel, 12-bit ADC (low- and high-gain) at a rate of 60 MSps. 
In addition, each module contains LED flashers for in-ice calibration, an FPGA for performing in-module local coincidence of PMT signals, and onboard $\mu$SD flash memory for buffering data before it is sent to the surface. In this contribution, we discuss the electronics and data acquisition system design.

\vspace{4mm}
{\bfseries Corresponding authors:}
Sean Griffin$^{1*}$ \\
{$^{1}$ \itshape WIPAC, University of Wisconsin-Madison, Madison, Wisconsin, USA}\\ [4mm]
$^*$ Presenter

\ConferenceLogo{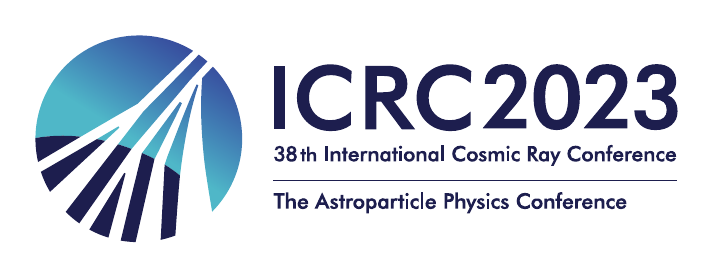}

\FullConference{The 38th International Cosmic Ray Conference (ICRC2023)\\ 26 July -- 3 August, 2023\\ Nagoya, Japan}
}

\begin{document}
\maketitle

\section{IceCube, IceCube Upgrade, and IceCube-Gen2}

The IceCube Neutrino Observatory \cite{IceCube} is a cubic kilometer of instrumented ice located at the geographic South Pole. 
IceCube-Gen2 \cite{IceCubeGen2} is a planned extension to IceCube which would increase the detection rate of cosmic neutrinos by an order of magnitude by adding an expansive optical array with $\sim$10,000 novel sensors, a radio array for extremely high energy (EHE) neutrinos, and a surface array for ultra high energy (UHE) cosmic rays.    
IceCube-Gen2 will be able to detect significantly fainter neutrino sources and measure neutrinos of energies several orders of magnitude greater than the current detector (colloquially referred to as Gen1).
The IceCube Upgrade \cite{IceCube-Upgrade} is a funded extension to IceCube and will be completed in early 2026.
As part of this upgrade, we are developing prototype modules in preparation for IceCube-Gen2. 


\section{Data Acquisition Architecture}

The primary detector in IceCube, the Digital Optical Module (DOM), measures Cherenkov photons produced by particle interactions in the ice. 
These ``hits'' are digitized and subsequently transmitted to the surface over a cable that supplies both power to the DOM and communications via a custom UART-over-power protocol.
In IceCube, all hit information is transmitted to the surface where computing nodes collect the information from each DOM and build events based on various trigger criteria. 
In Gen2, there is insufficient bandwidth available in the in-ice string to transmit all hit information to the surface. 
Instead, only timestamp information is transmitted to the surface, and the remaining hit information is buffered in-ice in nonvolatile memory (``the hitspool'').
If one of the trigger engines is satisfied, a request is made to each module to provide all data within a specified time window.

\section{IceCube-Gen2 DOM Concept}

\begin{figure}[hbtp]
     \centering
     \begin{subfigure}[b]{0.3\textwidth}
         \centering
         \includegraphics[height=\textwidth]{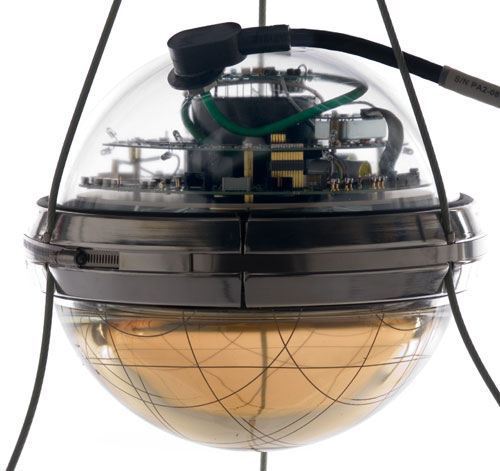}
         \caption{DOM \cite{Gen1-instr} (IceCube)}
     \end{subfigure}
     \hfill
     \begin{subfigure}[b]{0.3\textwidth}
         \centering
         \includegraphics[height=\textwidth]{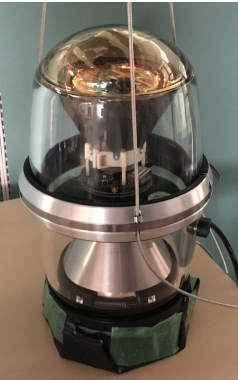}
         \caption{D-Egg \cite{degg} (IceCube Upgrade)}
     \end{subfigure}
    \hfill
     \begin{subfigure}[b]{0.3\textwidth}
         \centering
         \includegraphics[height=\textwidth]{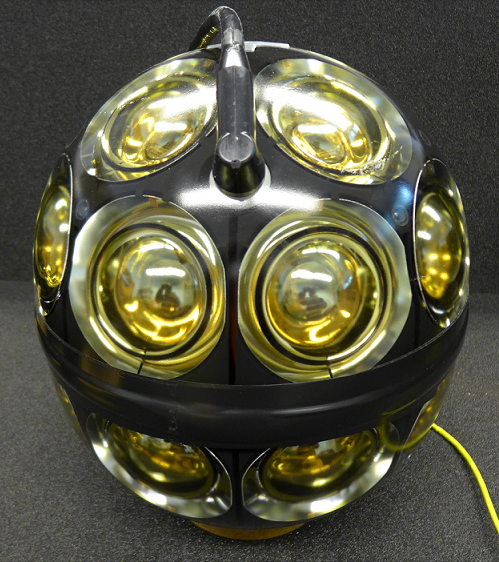}
         \caption{mDOM \cite{mdom} (IceCube Upgrade)}
     \end{subfigure}
     \vspace{1em}
     
      \begin{subfigure}[b]{0.6\textwidth}
         \centering
         \includegraphics[height=.65\textwidth]{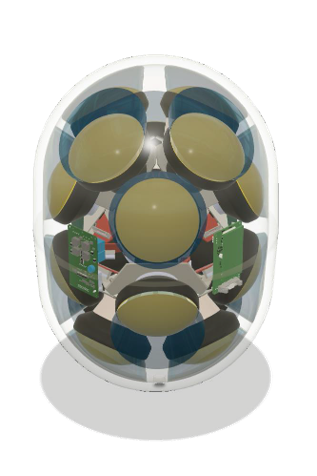}
         \includegraphics[height=.65\textwidth]{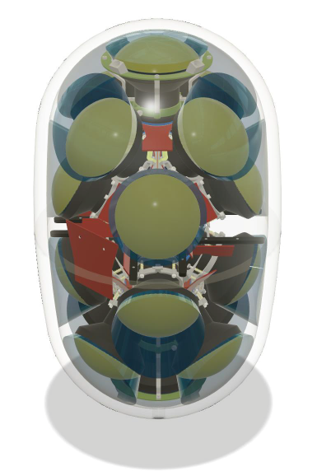}
         \caption{16-PMT (\textbf{left}) and 18-PMT (\textbf{right}) DOM (IceCube-Gen2)}
     \end{subfigure}
     \vspace{1em}
     \caption{
     The Gen1 DOM \textbf{(a)} has a single, downward-facing $\SI{10}{\inch}$ PMT.
    The D-Egg \textbf{(b)} comprises two \SI{8}{\inch} PMTs facing opposite directions. 
    The mDOM \textbf{(c)} has twenty-four \SI{3}{\inch} PMTs providing roughly uniform acceptance over all angles. 
    The IceCube-Gen2 DOM \textbf{(d)} will have 16 (or 18) $\SI{4}{\inch}$ PMTs. 
    }
    \label{fig:doms}
\end{figure}

The Gen2 DOM is designed to provide roughly uniform coverage over all angles, with directional information that is achieved by using multiple smaller PMTs covering two hemispheres. 
Conversely, the IceCube Gen1 DOM comprises one monolithic \SI{10}{\inch}, downward-facing PMT \cite{IceCubeGen1PMT} read out with a mainboard \cite{IceCubeGen1Mainboard}. 
Images of the IceCube DOM alongside two IceCube Upgrade modules and the Gen2 DOM concepts are given in \Cref{fig:doms}.
We have designed both a 16- and 18-PMT variant; the majority of the internal components are common to both. 
The size and shape of the Gen2 DOM prototype is the result of an optimization of cost, drilling time (which scales with the size of the borehole), and usable internal volume. 
Additional information on the Gen2 DOM mechanical structure can be found in Reference \cite{Gen2DOMICRC_Mechanical}.

\begin{figure}[hbtp]
    \centering
    \includegraphics[width=0.96\textwidth]{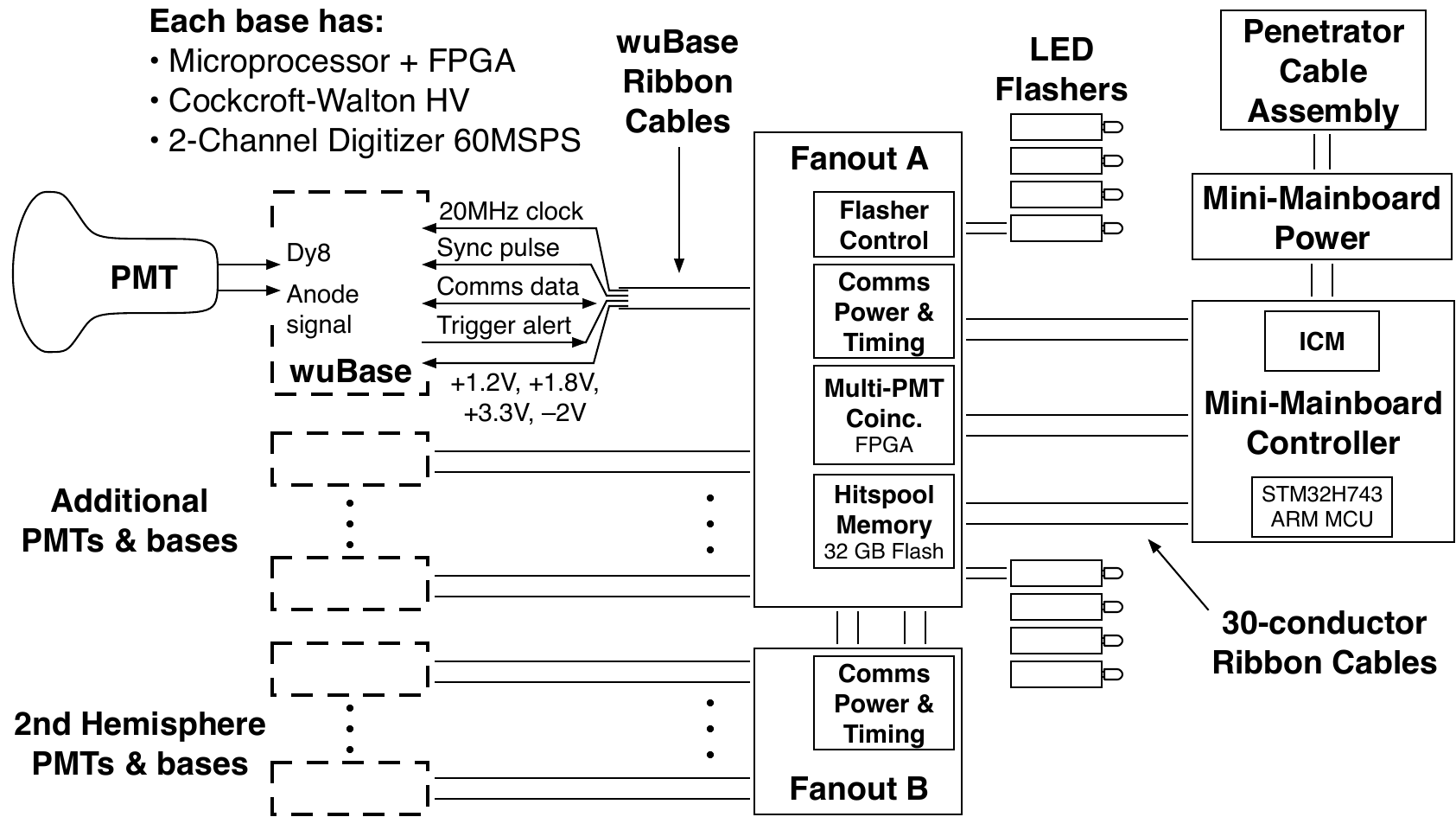}\\
    \vspace{1em}
    \includegraphics[width=0.96\textwidth]{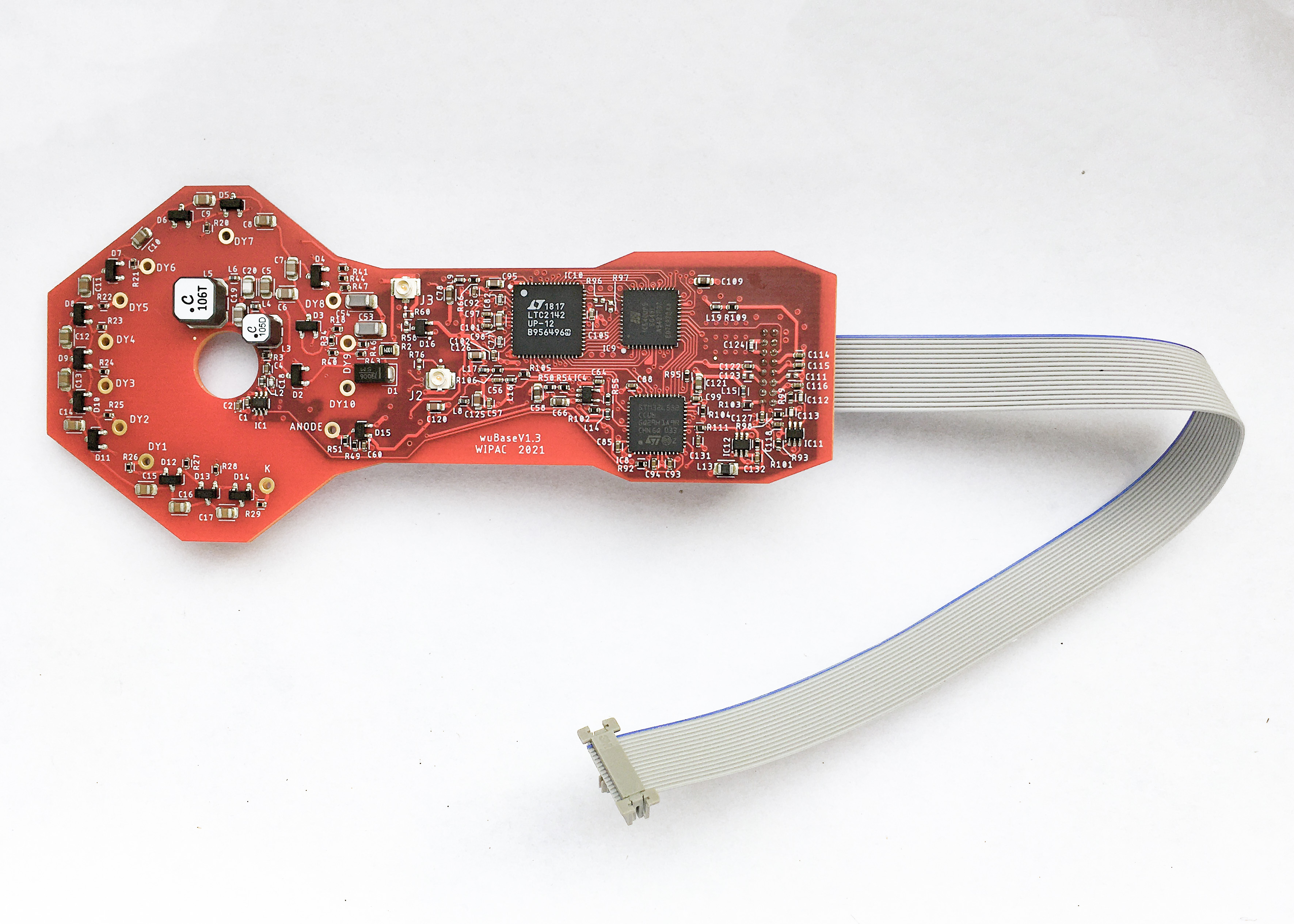}
    \caption{
    \textbf{Top:} Diagram of the Gen2 DOM dataflow.
    Cherenkov photons are detected by the PMT; the resulting waveform is digitized and buffered in the waveform microbase (wuBase).
    The mainboard MCU queries each wuBase for data via a UART multiplexer located on the fanout board and stores the hit data in onboard flash memory. 
    \textbf{Bottom:} Photo of a wuBase.
    }
    \label{fig:LOM-dataflow} 

\end{figure}

A block diagram of the main Gen2 DOM electrical components is given in the top panel of \Cref{fig:LOM-dataflow}. 
The prototype front-end electronics are based on the custom designed Waveform Microbase (wuBase, bottom panel of the same figure) and command and data handling (C\&DH) for the prototype DOM is handled by a ``Mini-Mainboard'' (MMB).  

\section{Electrical Design}
\subsection{Front-End}

\begin{figure}[hbtp]
    \centering
    \includegraphics[width=0.49\textwidth]{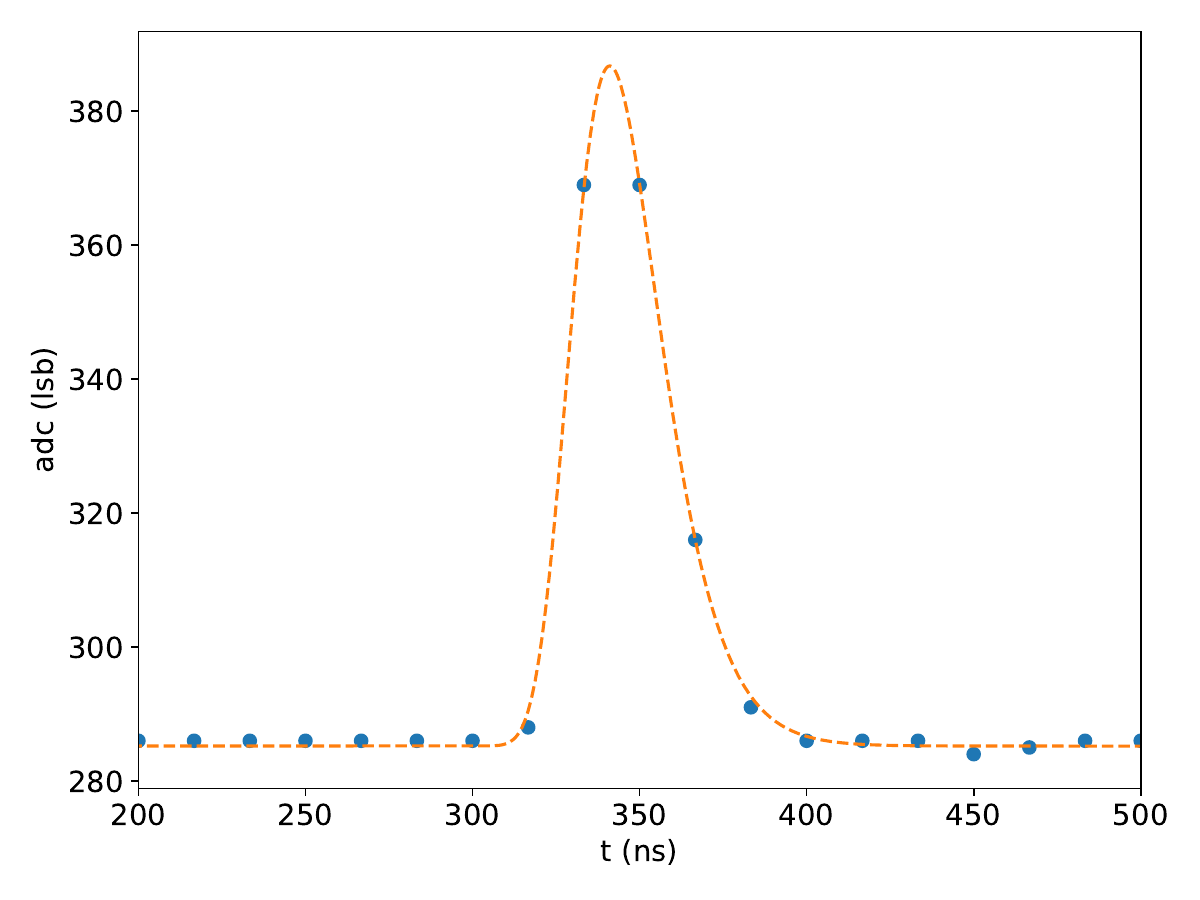}
    \includegraphics[width=0.49\textwidth]{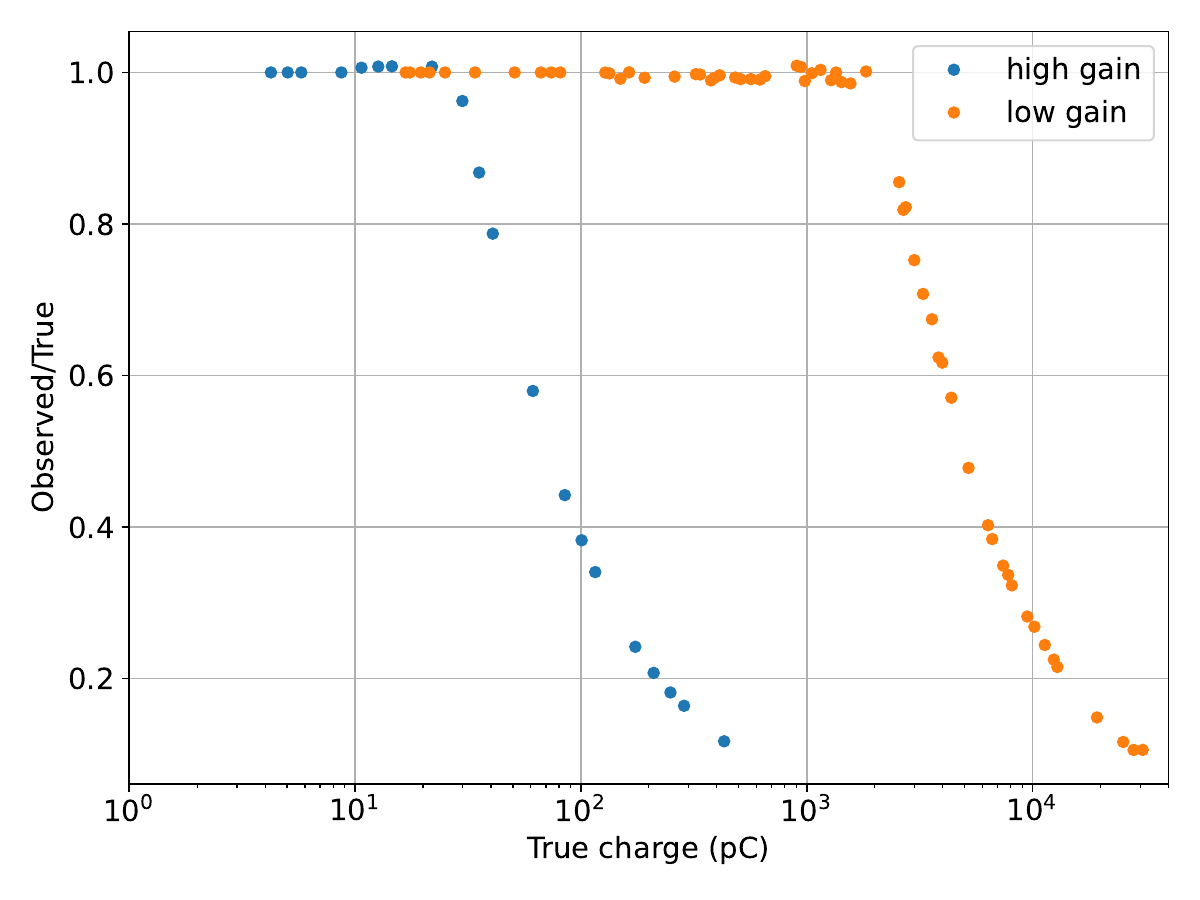}
    \includegraphics[width=0.98\textwidth]{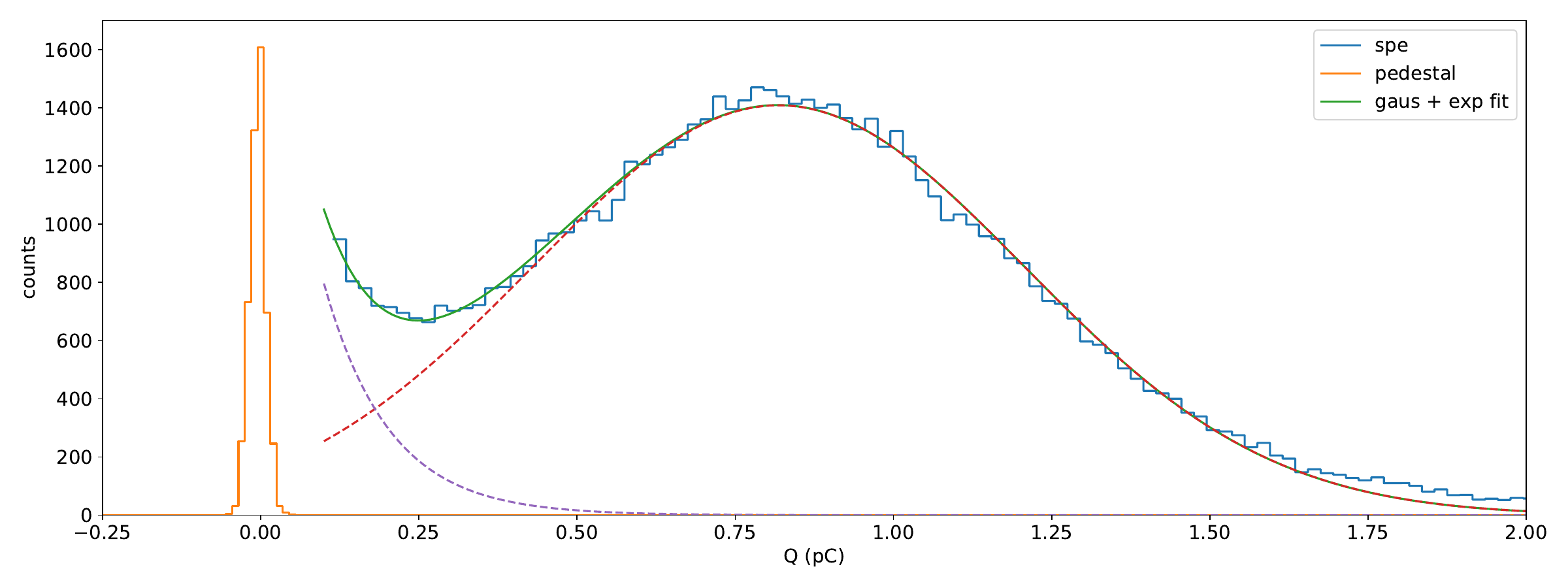}
    \caption{
        \textbf{Top left:} Example digitized wuBase pulse with a fit. 
        \textbf{Top right:} Demonstration of linearity for the two wuBase gain channels across a large dynamic range.
        \textbf{Bottom:} Typical SPE spectrum for a wuBase; the pedestal noise distribution is on the left (FWHM $\sim\SI{0.03}{\pico\coulomb}$), the SPE distribution FWHM is $\sim\SI{0.9}{\pico\coulomb}$. 
        The exponential and gaussian fitted components are also shown individually as dashed lines. }
    \label{fig:spe}     
\end{figure}

Each wuBase (1 per PMT) comprises a 2-channel digitizer operating at 60~Msps, providing 2 gain channels covering a wide and linear dynamic range (roughly 1-10,000 photoelectrons). 
The dynamic range requirement is driven by the need to be sensitive to both dim and bright Cherenkov events (\emph{i.e.} both low-energy (or distant) and high-energy (or close) physics events). 
The sampling rate has been chosen to maintain low power consumption for the device ($\lesssim \SI{150}{\milli\watt}$); a pulse shaping circuit has been finely tuned to allow PMT pulses to be fit and achieve a time resolution at the nanosecond level. 
A typical waveform and the corresponding pulse fit are given in \Cref{fig:spe} alongside an example linearity measurement for both gain channels and a single photoelectron (SPE) spectrum. 
Additional information on PMT performance for the Gen2 DOM prototypes can be found in Reference \cite{Gen2DOMICRC_PMT}. 

The ADCs are read out via a low-power FPGA (Lattice ICE40); data are buffered in one of two data ``pages'' until they are retrieved by the wuBase MCU (STM32L5) via SPI. 
The vast majority of ``hits'' are SPE events, which can be described entirely by a charge and timestamp. 
This allows data to be compressed significantly, minimizing the overall bandwidth required to send data to the surface. 
As such, in order to save bandwidth and maximize the depth of the in-ice hitspool, a simplified version of the IceCube ``WaveDeform'' algorithm (see Reference~\citenum{IceCube-erecon}) will be implemented on the wuBase MCU to classify hits as either SPEs or multiple photoelectron (MPE) waveforms (``complex waveforms''). 
Hits are buffered in the wuBase volatile memory until a request from the back-end electronics is received. 
The wuBase also generates a signal for each hit which is forwarded to an FPGA residing on a fanout board (described later); this FPGA will generate a coincidence flag based on these inputs. 

We have developed four variants of the wuBase to support the various Gen2 DOM configurations; two board shapes are required as the geometry is dictated by the spacing of the equatorial and polar PMTs, and two variants are required to support the two PMT vendors under consideration (Hamamatsu and NNVT).

\subsection{Back-End}

\subsubsection{Mini-Mainboard}

The Mini-Mainboard (MMB) acts as the command and data handling (C\&DH) processor for a number of IceCube Upgrade devices, including the Gen2 DOM prototype. 
It contains two primary components: an STM32H7 MCU for module C\&DH and an IceCube Communications Module (ICM, see references~\citenum{deggelectronics, mdomdesign}) which handles power distribution, timing, and surface communications.

The MMB itself is actually two boards: a processor board, containing the MCU, ICM, and I/O ports for various Upgrade modules; and a power board which handles the electrical interface with the main in-ice cable assembly, setting the hardware address on the device on the string, and generation of the main voltage rail for the processor board and subsequent peripherals. 

During operations, the MMB will query each wuBase and request available hits which will be stored in the hitspool. 
For the IceCube Upgrade, all hits will ultimately be sent to the surface, but in order to test all components of the Gen2-style dataflow, hits will still be buffered on the hitspool prior to being transmitted to the surface. 
Firmware and software in each FPGA and MCU is upgradeable after the devices have been deployed, with dual-redundant ``fallback'' firmware in the ICMs providing robustness against corrupted firmware images.

\subsubsection{Fanout Boards}

The Gen2 DOM prototype uses two fanout boards (one per hemisphere, referred to the A-side and B-side) to multiplex UART communications between the MMB and the various wuBases. 
In addition, the fanout boards provide the electrical interface to LED flasher modules which are used for in-ice calibration.
The A-side fanout board will host two $\mu$SD cards to support the hitspool.
The two SD cards can be individually addressed via a demultiplexer chip.
An additional Lattice ICE40 FPGA resides on the A-side fanout board to handle the generation of PMT coincidence signals. 
This coincidence signal reduces the overall data rate: hits seen simultaneously in multiple PMTs are more likely to be due to physics events than PMT dark noise. 
This coincidence information is sent to the surface and will be used as part of the array trigger.

\subsubsection{Hitspool}

Bandwidth testing has been performed to ensure that the flash storage can keep up with the PMT data rates. 
The estimated data rate for a Gen2 DOM is roughly \SI{500}{\kilo\bit/\second} after data compression.
We have tested the SDMMC interface\footnote{Details on the hardware interface and communications standard can be found in Reference \citenum{sdmmc}.} on the STM32H7 MCU; the read/write bandwidth (\Cref{tab:hitspool_bandwidth}) exceeds the requirement by a significant margin. 
Note that due to the overhead involved in I/O operations, the read and write ``block sizes'' affect the speed which the MCU can access the hitspool. 
The choice of block size will therefore require optimization of the number of available compute cycles on the MCU and the amount of memory allocated to data buffering. 

\begin{table}[hbtp]
\begin{center}
\begin{tabular}{r|c|c}
\hline\hline
Block Size (B) & \multicolumn{2}{c}{Bandwidth (MB/s)} \\ \hline
               & Read             & Write             \\ \hline
0x1000         & 2.61             & 0.96              \\
0x5000         & 4.42             & 1.64              \\
0x10000        & 5.09             & 2.97              \\\hline\hline
\end{tabular}
\end{center}
\caption{Example mini-mainboard $\mu$SD bandwidth measurements.}
\label{tab:hitspool_bandwidth}
\end{table}

We have selected the Swissbit \SI{32}{\giga\byte} S-56u $\mu$SD card for the Gen2 DOM prototypes. 
The depth of the hitspool (\SI{32}{\giga\byte}) is based on the Gen2 requirement to buffer data for $\sim 1~\mathrm{week}$ to support sub-threshold searches for Galactic supernovae in IceCube data (see Reference~\citenum{icecube-supernovae-sensitivity}).
The card uses pseudo-SLC flash memory, is rated to operate and retain data from \SIrange[mode=text,range-phrase = {\text{~to~}}]{-40}{85}{\celsius}, covering the expected operation regime of the in-ice modules (typically \SIrange[mode=text,range-phrase = {\text{~to~}}]{-32}{-8}{\celsius}). 
In addition, through private communications with Swissbit, we expect the devices to continue to function as expected down to \SI{-60}{\celsius}. 
The devices contain on-board error correction code (ECC) with a mean time between failures of $>$\SI{3000000}{hours}. 
In addition, libraries are available to read internal hardware registers to monitor device health (similar to SMART data for solid state and hard drives). 

Data are stored in a FAT filesystem \cite{fatfs}; to minimize file I/O overhead, data arrive from the wuBases pre-formatted to save on MCU cycles. 
In addition, file sizes will be optimized to minimize overhead and search times. 
To provide robustness against power failures which reset the nonvolatile memory in the MMB MCU, the hitspool search index is encoded in the directory structure: file and folder names are mapped to timestamps, so in the event that the MCU is reset, the index can be reconstructed by scanning the filesystem rather than needing to scan the contents of hitspool files themselves. 

\section{Outlook}

The development of the first Gen2 DOM prototype is at a high state of completion. 
A mechanical model comprising all major components (albeit with dummy PMTs and electronics) was completed in June 2023 and an electronic prototype hemisphere will be completed in early Fall 2023 with the objective of a completed module by the end of the year. 

Twenty prototype modules will be constructed, with twelve ultimately deployed in the ice as part of IceCube Upgrade. 
Both the 16- and 18-PMT DOM prototypes will be deployed as a proof-of-concept in preparation for IceCube-Gen2, ultimately informing the design for the upcoming IceCube-Gen2 proposal.

\bibliography{bib} 



\clearpage
\section*{Full Author List: IceCube-Gen2 Collaboration}

\scriptsize
\noindent
R. Abbasi$^{17}$,
M. Ackermann$^{76}$,
J. Adams$^{22}$,
S. K. Agarwalla$^{47,\: 77}$,
J. A. Aguilar$^{12}$,
M. Ahlers$^{26}$,
J.M. Alameddine$^{27}$,
N. M. Amin$^{53}$,
K. Andeen$^{50}$,
G. Anton$^{30}$
% ...
\\
\\
% ...
$^{12}$ Universit{\'e} Libre de Bruxelles, Science Faculty CP230, B-1050 Brussels, Belgium \\
% ...
$^{17}$ Department of Physics, Loyola University Chicago, Chicago, IL 60660, USA \\
% ...
$^{22}$ Dept. of Physics and Astronomy, University of Canterbury, Private Bag 4800, Christchurch, New Zealand \\
% ...
$^{26}$ Niels Bohr Institute, University of Copenhagen, DK-2100 Copenhagen, Denmark \\
$^{27}$ Dept. of Physics, TU Dortmund University, D-44221 Dortmund, Germany \\
% ...
$^{30}$ Erlangen Centre for Astroparticle Physics, Friedrich-Alexander-Universit{\"a}t Erlangen-N{\"u}rnberg, D-91058 Erlangen, Germany \\
% ...
$^{47}$ Dept. of Physics and Wisconsin IceCube Particle Astrophysics Center, University of Wisconsin{\textendash}Madison, Madison, WI 53706, USA \\
% ...
$^{50}$ Department of Physics, Marquette University, Milwaukee, WI, 53201, USA \\
% ...
$^{53}$ Bartol Research Institute and Dept. of Physics and Astronomy, University of Delaware, Newark, DE 19716, USA \\
% ...
$^{76}$ Deutsches Elektronen-Synchrotron DESY, Platanenallee 6, 15738 Zeuthen, Germany  \\
$^{77}$ Institute of Physics, Sachivalaya Marg, Sainik School Post, Bhubaneswar 751005, India \\
% ...

\subsection*{Acknowledgements}

\noindent
The authors gratefully acknowledge the support from the following agencies and institutions:
USA {\textendash} U.S. National Science Foundation-Office of Polar Programs,
U.S. National Science Foundation-Physics Division,
U.S. National Science Foundation-EPSCoR,
Wisconsin Alumni Research Foundation,
Center for High Throughput Computing (CHTC) at the University of Wisconsin{\textendash}Madison,
Open Science Grid (OSG),
Advanced Cyberinfrastructure Coordination Ecosystem: Services {\&} Support (ACCESS),
Frontera computing project at the Texas Advanced Computing Center,
U.S. Department of Energy-National Energy Research Scientific Computing Center,
Particle astrophysics research computing center at the University of Maryland,
Institute for Cyber-Enabled Research at Michigan State University,
and Astroparticle physics computational facility at Marquette University;
Belgium {\textendash} Funds for Scientific Research (FRS-FNRS and FWO),
FWO Odysseus and Big Science programmes,
and Belgian Federal Science Policy Office (Belspo);
Germany {\textendash} Bundesministerium f{\"u}r Bildung und Forschung (BMBF),
Deutsche Forschungsgemeinschaft (DFG),
Helmholtz Alliance for Astroparticle Physics (HAP),
Initiative and Networking Fund of the Helmholtz Association,
Deutsches Elektronen Synchrotron (DESY),
and High Performance Computing cluster of the RWTH Aachen;
Sweden {\textendash} Swedish Research Council,
Swedish Polar Research Secretariat,
Swedish National Infrastructure for Computing (SNIC),
and Knut and Alice Wallenberg Foundation;
European Union {\textendash} EGI Advanced Computing for research;
Australia {\textendash} Australian Research Council;
Canada {\textendash} Natural Sciences and Engineering Research Council of Canada,
Calcul Qu{\'e}bec, Compute Ontario, Canada Foundation for Innovation, WestGrid, and Compute Canada;
Denmark {\textendash} Villum Fonden, Carlsberg Foundation, and European Commission;
New Zealand {\textendash} Marsden Fund;
Japan {\textendash} Japan Society for Promotion of Science (JSPS)
and Institute for Global Prominent Research (IGPR) of Chiba University;
Korea {\textendash} National Research Foundation of Korea (NRF);
Switzerland {\textendash} Swiss National Science Foundation (SNSF);
United Kingdom {\textendash} Department of Physics, University of Oxford.

\end{document}